\documentstyle [12pt]{article}
\advance\voffset by -1 in
\advance\hoffset by -.5 in
\begin{document}
\def\p{\partial}
\def\a{\alpha}
\def\b{\beta}
\def\g{\gamma}
\def\tr{{\rm tr}\:}
\def\res{{\rm res}\:}
\def\tres{\tr\res}
\def\cA{{\cal A}}
\def\l{\lambda}
\def\L{\Lambda}
\def\f{\phi}
\def\hw{\hat{w}}
\def\hv{\hat{v}}
\def\t{\tau}
\def\Tau{T}
\def\rarr{\rightarrow}
\def\bg{{\bf g}}
\def\d{\delta}
\def\s{\sum}
\def\op{^{\oplus}}
\def\om{^{\ominus}}
\def\tl{\tilde{\lambda}}
\def\bof{{\bf f}}
\def\bb{{\bf b}}
\def\bh{{\bf h}}
\def\bc{{\bf c}}
\def\be{{\bf e}}
\def\F{\Phi}
\centerline{{\bf WHY THE GENERAL ZAKHAROV-SHABAT EQUATIONS }}
\centerline{{\bf FORM A HIERARCHY?}}
\vspace{.5in}
\centerline{{\bf L.A.Dickey}}
\centerline{Dept. Math., University of Oklahoma, Norman, OK 73019}
\centerline{Internet: ldickey@nsfuvax.math.uoknor.edu}

\vspace{.3in}
\begin{abstract}
The totality of all Zakharov-Shabat equations (ZS), i.e., zero-curvature
equations with rational dependence on a spectral parameter, if properly
defined, can be considered as a hierarchy. The latter means a collection
of commuting vector fields in the same phase space. Further properties
of the hierarchy are discussed, such as additional symmetries, an analogue
to the string equation, a Grassmannian related to the ZS hierarchy, and a
Grassmannian definition of soliton solutions.
\end{abstract}

\vspace{.3in}
{\bf 0. Introduction.}
We are accustomed to the fact that integrable systems appear not one at a time
but in big families called hierarchies. So, first of all, the KdV ($n=2$)
hierarchy was invented (Gardner, Green, Kruskal, and Miura made the first and
the most important discovery, the KdV equation in the proper sense; later on
all the higher KdV were found by Gardner). Then this was generalized to every $
n$ (Gelfand and Dickey, who used fractional powers of operators). Thus,
infinitely many generalized KdV hierarchies were found. They were unified to
a single one large KP
hierarchy (Kyoto school: Sato et al. [1]). Another line of developments was
connected with equations generated by a linear first order differential
operator with matrix coefficients linearly dependent on a spectral parameter
(Ablowitz, Kaup, Newell and Segur for matrices $2\times 2$ and Dubrovin in a
general case, let us call this hierarchy AKNS-D). Later this was generalized to
operators with a polynomial dependence on the spectral parameter (Dickey,
Reyman and Semenov-Tjan-Shanski). Thus, for every degree of a polynomial, $m$,
there is a hierarchy, a generalized AKNS-D.\footnote{More detail can be found,
e.g., in [3].}

More than that, there is a very general type of equations proposed by Zakharov
and Shabat (see [2]): the equation of zero curvature, where matrices
depend on some spectral parameter as rational functions. (The above mentioned
hierarchies where operators depend on a parameter as polynomials, i.e.,
have a single pole at infinity, represent special case of these equations).
The ZS equations usually have been treated individually, not as a hierarchy.

Which property permits to consider a hierarchy as a single whole, as an entity?
Geometrically speaking, a differential equation is a flow in a phase space.
We call a family of differential equations a hierarchy if they act in the same
phase space and commute. Then each of equations determines a symmetry for each
other. Let us consider this as a characteristic property of a hierarchy.

We see our goal in this paper in the construction of a theory where all the
Zakharov-Shabat (ZS) equations can be considered as one hierarchy, in the above
sense.

We discuss also problems such as a Grassmannian approach to the ZS hierarchy,
the existence of additional symmetries and a definition of an analogue to
the string equation.  \\

The study of integrable systems is interesting in several aspects. One is
a desire to expand the class of exactly solvable equations or to find new
solutions to already known integrable equations. This can be important in
applications, though one must realize that there are rather few integrable
systems among all equations significant for physics or engineering. The other
aspect is that this area of mathematics provides very rich algebraic structures
which more and more find their way to modern physical theories. The fact that
there are large hierarchies of commuting equations is of great importance just
from this point of view.\\

{\bf 1. Definition of the ZS hierarchy.} Let $a_k$, $k=1,...,m$ be a given set
of complex numbers. Let, for every $k$, $$\hw_
k=\sum_0^\infty w_{ki}(z-a_k)^i,$$ be a formal series, the entries of matrices
$w_{ki}$ being taken as generators of a differential
algebra. Then this algebra is extended by elements $(\det w_{k0})^{-1}$ and the
obtained algebra is called $\cA_w$. The formal series $\hw_k$ can be inverted
within this algebra. Let $$R_{k\a}=\hw_kE_{\a}\hw_k^{-1};~~R_{k\a l}=R_{k\a}(z
-a_k)^{-l}$$ where $E_\a$ is a matrix with only one non-vanishing element,
equal
1, on the $(\a,\a)$ place.

We consider two kinds of objects. Such quantities as $\hw_k$ and $R_{k\a l}$
are formal series, or jets, at the points $a_k$. The algebra of all such jets
will be called $J_k$ and $J=\oplus J_k$. If $j_k\in J_k$ is a
jet then $j_k^-$ symbolizes its principal part, i.e., a sum of negative
powers of $z-a_k$, and $j_k^+$ the rest of the series. If the principal part
contains finite number of terms (and we tacitly assume this unless the opposite
is said or is evident from a context) it can be considered as a global
meromorphic function; the algebra of global meromorphic functions is $G$.
Global functions are objects of the second kind. A global meromorphic function
gives rise to a jet at every $a_k$. In particular, $j_k^-$ can be
considered as a jet at a point $a_{k_1}$, different from $a_k$, more precisely,
as an element of $J_{k_1}^+$.\\

{\bf Definitions. 1.} {\sl A hierarchy corresponding to a fixed set $\{a_k\}$
is defined by the equations
$$\p_{k\a l}\hw_{k_1}=\left\{\begin{array}{l}-R_{k\a l}^+\hw_{k_1},~~k=k_1\\
{}~~R_{k\a l}^-\hw_{k_1},~~{\rm otherwise}\end{array}\right.,~~~\p_{k\a
l}=\p/\p t_{k\a l}.\eqno{(1)}$$ In the second case $R_{k\a l}^-$ is
considered as an element of $J_{k_1}^+$, see above; $t_{k\a l}$ are some
variables.}\\

{\bf 2.} {\sl A ZS hierarchy is an inductive limit of hierarchies with fixed
sets $\{a_k\}$, with respect to a natural embedding of a hierarchy
corresponding
to a subset into a hierarchy corresponding to a set, as a subhierarchy.}\\

Further in this section, for simplicity of writing, we shall unite indices $\a$
and $l$ into one subscript $a=(\a,l)$ and write $\p_{ka}$ and $R_{ka}$ instead
of $\p_{k\a l}$ and $R_{k\a l}$.

(It would be possible to take in the above definitions at each point $a_k$ its
own spectral family $E_{k\a}$, $\a=1,..,n$, which not necessarily commute for
distinct $k$; however, it is easy to see that this is not a genuine
generalization since it can be reduced to the same by a substitution $\hw_k
\mapsto\hw_kc_k^{-1}$ where $c_k$ are matrices reducing $E_{k\a}$ to the
diagonal form, $E_{k\a}=c_kE_\a c_k^{-1}$.)\\

{\bf Lemma.} {\sl Equalities $$\p_{k\a l}R_{k_1\a_1l_1}=\left\{\begin{array}{l}
-[R_{k\a l}^+,R_{k_1\a_1l_1}],~~k=k_1\\~~[R_{k\a l}^-,R_{k_1\a_1l_1}],~~{\rm
otherwise}\end{array}\right.$$ hold.}\\

{\em Proof.} It easily can be obtained from the definition of $R_{k\a l}$.
$\Box$  \\

{\bf Theorem.} {\sl All operators $\p_{k\a l}$ commute.}\\

{\em Proof.} One has to prove $[\p_{k_1a_1},\p_{k_2a_2}]\hw_{k_3}=0$ in 3
cases: i) all of $k_i$ coincide, ii) only two of them coincide, iii) all are
distinct.

i)  We have $$(\p_{ka_1}\p_{ka_2}-\p_{ka_2}\p_{ka_1})\hw_k=-\p_{ka_1}R_{ka_2}^+
\hw_k-(1\Leftrightarrow 2)$$ $$=[R_{ka_1}^+,R_{ka_2}]_k^+\hw_k+R_{ka_2}^+R_{ka_
1}^+\hw_k-(1\Leftrightarrow 2)=R_{ka_1}^+R_{ka_2}^+\hw_k+[R_{ka_1},R_{ka_2}^-]_
k^+\hw_k$$ $$-[R_{ka_2}^+,R_{ka_1}]_k^+\hw_k-R_{ka_1}^+R_{ka_2}^+\hw_k=[R_{ka_1
},R_{ka_2}]_k^+\hw_k=0$$since $R_{ka_1}$ and $R_{ka_2}$ commute. A notation $A_
k^-$ means the principal part of an expansion of $A$ in powers of $z-a_k$.
Similarly, $A_k^+$.

ii) First we consider $$(\p_{ka_1}\p_{ka_2}-\p_{ka_2}\p_{ka_1})\hw_{k_1}
=\p_{ka_1}R_{ka_2}^-\hw_{k_1}-\p_{ka_2}R_{ka_1}^-\hw_{k_1}$$ $$=-[R_{ka_1}^+,R_
{ka_2}]_k^-\hw_{k_1}+R_{ka_2}^-R_{ka_1}^-\hw_{k_1}-(1\Leftrightarrow 2)$$ $$=-(
R_{ka_1}^+R_{ka_2})_k^-\hw_{k_1}+(R_{ka_2}^-R_{ka_1})_k^-\hw_{k_1}-(1
\Leftrightarrow 2)=-[R_{ka_1},R_{ka_2}]\hw_{k_1}$$ which is zero since $R_{ka_1
}$ and $R_{ka_2}$ with the same $k$ commute.

Then we take $$(\p_{ka}\p_{k_1a_1}-\p_{k_1a_1}\p_{ka})\hw_{k}=
\p_{ka}R_{k_1a_1}^-\hw_{k}+\p_{k_1a_1}R_{ka}^+\hw_{k}$$ $$=
[R_{ka}^-,R_{k_1a_1}]_{k_1}^-\hw_{k}-R_{k_1a_1}^-R_{ka}^+\hw_{k}
+[R_{k_1a_1}^-,R_{ka}]_k^+\hw_{k}+R_{ka}^+R_{k_1a_1}^-
\hw_{k}$$ $$=[R_{ka}^-,R_{k_1a_1}]_{k_1}^-\hw_{k}-
[R_{k_1a_1}^-,R_{ka}^+]\hw_k+[R_{k_1a_1}^-,R_{ka}^+]_k^
+\hw_{k}+[R_{k_1a_1}^-,R_{ka}^-]_k^+\hw_{k}$$
$$=([R_{ka}^-,R_{k_1a_1}^-]_{k_1}^--[R_{ka}^-,R_{k_1a_1}^-]+[(R_{ka})_k^-,R_{k_
1a_1}^-]_k^-)\hw_k.$$ In the parentheses
there is a function $[R_{ka}^-,R_{k_1a_1}^-]$ minus its principal
parts in both the poles, $a_k$ and $a_{k_1}$. Thus, this is a constant. This
expression approaches zero when $z\rightarrow\infty$ which implies that the
constant is zero.

iii)$$[\p_{k_1a_1},\p_{k_2a_2}]\hw_{k_3}=\p_{k_1a_1}R_{k_2a_2}^-\hw_{k_
3}-\p_{k_2a_2}R_{k_1a_1}^-\hw_{k_3}$$ $$=[R_{k_1a_1}^-,R_{k_2a_
2}^-]_{k_2}^-\hw_{k_3}+R_{k_2a_2}^-R_{k_1a_1}^-\hw_{k_3}
-(1\Leftrightarrow 2)$$ $$=([R_{k_1a_1}^-,R_{k_2a_2}^-]_{k_1}^-
+[R_{k_1a_1}^-,R_{k_2a_2}^-]_{k_2}^--[R_{k_1a_1}^-,R_{k_2a_2}^-])\hw_{k_3}.$$
The expression in the parentheses vanish by
the same reason as in the previous case. $\Box$\\

{\bf 2. Gauge.} If we let $\hv_k=c\hw_k$ where $c$ is a matrix depending of
variables $\{t_{k\a l}\}$ in an arbitrary way, then $\hv_k$ satisfy
equations $$\p_{k\a l}\hv_{k_1}=A_{k\a l}\hv_{k_1}+\left\{\begin
{array}{l}-R_{k\a l}^+\hv_{k_1},~~k=k_1\\~~R_{k\a l}^-\hv_{k_1},~~{\rm
otherwise}\end{array}\right.\eqno{(2)}$$ with new $R_{k\a l}=\hv_kE_\a(z-a_k)
^{-l}\hv_k^{-1}$, and $A_{k\a l}=\p_{k\a l}c\cdot c^{-
1}$ which implies $$\p_{k_1\a_1l_1}A_{k\a l}-\p_{k\a l}A_{k_1\a_1l_1}+
[A_{k\a l},A_{k_1\a_1l_1}]=0.\eqno{(3)}$$
These equations are slightly more general than (1). We say that $\hw_k$ and
$\hv_k$ are gauge-equivalent. Conversely, if some $\hv_k$ satisfy (2) with
the property (3) for $A$, then, integrating equations $\p_{k\a l}c=A_{k\a
l}c$ (which are compatible by virtue of this property), one can find a
gauge-equivalent $\hw_k$ satisfying (1). We can, e.g., normalize a solution by
a condition $\sum_i\hw_i(a_i)=I$. The following lemma can be useful:\\

{\bf Lemma.} {\sl Given solutions of (2) with some $A_{k\a l}$ not depending on
$z$ for any $(k\a l)$, being $\sum_i\hat{v}_i(a_i)=I$. Then there is some $c$
such that $A_{k\a l}=\p_{k\a l}c\cdot c^{-1}$ and, therefore, $\hv=\{\hv_i\}$
is gauge equivalent to a solution of (1).}\\

{\em Proof}. First of all from the assumption we get $$A_{k\a l}=R_{k\a
l}^+(a_k)\hv_k(a_k)-\sum_{k_1\neq k}R_{k\a l}^-(a_{k_1})\hv_{k_1}(a_{k_1}),$$
thus, $A_{k\a l}$ is a differential polynomial in elements of $\hv_i$'s. A new
differentiaton
can be defined in $\cA_v$: $\p_{k\a l}^*\hv_i=(\p_{k\a l}-A_{k\a l})\hv_i$.
The quantities $\hv_k$ satisfy the system (1) with respect to new variables
$t_{k\a l}^*$ and, therefore, $\p_{k\a l}^*$ commute, i.e., $\p_{k\a l}-A_{k\a
l}$ commute which is equivalent to Eq.(3). The rest is clear. $\Box$\\

Functions $\hw_k$ admit also the following transformations: multiplication
on the right by series in $(z-a_k)^{-1}$ with constant diagonal coefficients.
This does not affect the equations (1) (or (2)) at all.\\

{\bf 3. Differential operators.} The following proposition readily can be
proven by a simple straightforward computation:\\

{\bf Proposition 1.} {\sl A dressing formula $$ \hw_{k_1}(\p_{k\a l}-E_{\a}
(z-a_k)^{-l}\d_{kk_1})\hw_{k_1}^{-1}=\p_{k\a l}-B_{k\a l},~~B_{k\a l}=
R_{k\a l}^-\eqno{(4)}$$ holds, as a consequence of Eq.(1).}\\

The operator $\p_{k\a l}-B_{k\a l}$ is assumed to act in $J_{k_1}$. However, it
does not depend on $k_1$ at all and can be considered as a global function of
$z$ with the only pole of the $l$th order at $a_k$.

Let $$w_k=\hw_k\exp\xi_k~~{\rm where}~\xi_k=\sum_{l=0}^\infty\sum_{\a=1}^nt_
{k\a l}E_{\a}(z-a_k)^{-l}.$$

{\bf Definition.} {\sl The collection $w=\{w_k\}$ is the formal Baker function
of the hierarchy.} \\

Eq.(4) can be rewritten in terms of the Baker function as $$w_{k_1}\p_{k\a l}
w_{k_1}^{-1}=\p_{k\a l}-B_{k\a l}.\eqno{(5)}$$

{\bf Proposition 2.} {\sl All the operators $\p_{k\a l}-B_{k\a l}$ commute.}\\

{\em Proof.} This is a corollary of the theorem Sect.1. and Eq.(5).
$\Box$\\

One can consider arbitrary linear combinations of the above constructed
operators,$$ L=\sum_{k,\a,l}\l_{k\a l}(\p_{k\a l}-B_{k\a l})=\p+U,\eqno{(6)}$$
where $\p=\sum_{k,\a,l}\l_{k\a l}\p_{k\a l}$ and $U=-\sum_{k,\a,l}\l_{k\a l}B_
{k\a l}$. Two such operators commute which yields equations of the
Zakharov-Shabat type $$\p U_1-\p_1 U=[U_1,U].$$ Functions $U$ and $U_1$ are
rational functions of the parameter $z$.\\

{\bf Remark 1.} {\sl It is possible to give a group theory interpretation to
these equation considering a Lie algebra of jets $J$ and, as the dual space,
global meromorphic functions with poles at $\{a_k\}$.} (See also [8]).\\

{\bf Remark 2.} {\sl Here we have a special case of ZS equation: the functions
$U$ and $U_1$ vanishing at infinity. If we make a gauge transformation $w_k
\mapsto cw_k$ then $\p+U\mapsto c(\p_{k\a l}+U)c^{-1}=\p_{k\a l}+cUc^{-1}-
(\p_{k\a l}c)c^{-1}$, the last term does not vanish at infinity. This yields
the general case.}\\

{\bf 4. Additional symmetry and the string equation.} It is well-known ([5])
that the KP hierarchy has infinitely many symmetries that are not contained in
the hierarchy itself; their characteristic feature is an explicit dependence on
the variables $t$. They are called ``additional symmetries". The so-called
``string" equation is nothing but a condition that our operators do not depend
on a parameter of an additional symmetry ([6]). We are going to suggest an
additional symmetry and the corresponding string equation for the ZS hierarchy.

Let $$\p_z-M_i=w_i\p_zw_i^{-1}=\p_z-\p_zw_i\cdot w_i^{-1}=\p_z-\p_z\hw_i\cdot
\hw_i^{-1}-\hw_i\xi_{iz}\hw_i^{-1}$$ where $\xi_{iz}=\p\xi/\p
z=-\s_a\s_{l=1}^\infty t_{i\a l}E_\a l(z-a_i)^{-l-1}$. The quantity $M_i=
\p_z\hw_i\cdot\hw_i^{-1}+\hw_i\xi_{iz}\hw_i^{-1}$ is a jet at the point $a_i$.

Dressing an obvious relation $[\p_z,\p_{k\a l}]=0$ with the help of $w_i$ at
the point $a_i$ gives $$[\p_z-M_i,\p_{k\a l}-B_{k\a l}]=0,$$ i.e.,
$\p_{k\a l}M_i=\p_zB_{k\a l}-[M_k,B_{k\a l}]$. Taking negative and positive
parts, we get at the point $a_i$ $$\begin{array}{llll}(1)~i=k~~&\p_{k\a
l}M_k^-&
=\p_zB_{k\a l}&-[M_k,B_{k\a l}]_k^-\\ &\p_{k\a l}M_k^+&=&-[M_k,B_{k\a l}]_k^+\\
(2)~i\neq k~~&\p_{k\a l}M_i^-&=&-[M_i,B_{k\a l}]_i^-\\ &\p_{k\a l}M_i^+&=\p_z
B_{k\a l}&-[M_i,B_{k\a l}]_i^+\end{array}\eqno{(7)}$$

{\bf Definition.} {\sl The additional symmetry is given by the system of
differential equations}
$$\p^*\hw_j=(-M_j^++\sum_{i\neq j}M_i^-)\hw_j.$$ The same equation can be
written also as $$\p^*\hw_j=-\p_z\hw_j-\s_i\s_\a\s_lt_{i\a
l}l\p_{i\a(l+1)}\hw_j
.$$ Formally, there are infinitely many terms in this series. It is possible to
freeze all $t_{i\a l}$ as zero, except for finite number of them.

As it is easy to see, the equation of the additional symmetry
implies $$\p^*R_{k\a l}=[-M_k^++\sum_{i\neq k}M_i^-,R_{k\a l}].\eqno{(8)}$$
This is an equality in $J_k$.\\

{\bf Proposition.} {\sl The additional symmetry commutes with operators of the
hierarchy, $[\p^*,\p_{k\a l}]=0$, i.e., it is a symmetry, indeed.}\\

{\em Proof}. We have to prove that for all $j$ a relation $[\p^*,\p_{k\a
l}]\hw_
j=0$ holds. We consider two cases. (1) $k\neq j$.
$$\p^*\p_{k\a l}\hw_j-\p_{k\a l}\p^*\hw_j=\p^*B_{k\a l}\hw_j-\p_{k\a l}(-M_j^+
+\s_{i\neq j}M_i^-)\hw_j$$ $$=[-M_k^++\s_{i\neq k}M_i^-,B_{k\a l}]_k^-\hw_j+
[B_{k\a l},-M_j^++\s_{i\neq j}M_i^-]_j\hw_j$$ $$+(\p_zB_{k\a l})\hw_j-[M_j,B_{k
\a l}]_j^+\hw_j+\s_{i\neq j,k}[M_i^-,B_{k\a l}]_i^-\hw_j-(\p_zB_{k\a l})\hw_j
+[M_k,B_{k\a l}]_k^-\hw_j.$$  First, discuss the
terms with $i\neq k,j$. These are $$\s_{i\neq k,j}[M_i^-,B_{k\a l}]_k^-
-\s_{i\neq k,j}[M_i^-,B_{k\a l}]+\s_{i\neq k,j}[M_i^-,B_{k\a l}]_i^-.$$
This is a difference between a global function $\s_{i\neq k,j}[M_i^-,B_{k\a
l}]$
and all its principal parts, at $a_k$ and $a_i$; it has to be constant. Taking
into account that it vanishes at infinity, we can conclude that this is zero.
It remains to calculate $$(-[M_k^+,B_{k\a l}]_k^-+[M_j^-,B_{k\a l}]_k^-+
[M_j^+,B_{k\a l}]_j-[M_k^-,B_{k\a l}]_j-[M_j,B_{k\a l}]_j^++[M_k,B_{k\a l}]_k
^-)\hw_j.$$ Notice that in the fourth term the subscript $j$ can be skipped,
this is a global term. The first, the fourth and the sixth terms cancel out.
The
remaining terms are $$[M_j^-,B_{k\a l}]_k^-+[M_j^+,B_{k\a l}]_j-[M_j,B_{k\a l}]
_j^+=[M_j^-,B_{k\a l}]_k^-+[M_j^+,B_{k\a l}]_j^--[M_j^-,B_{k\a l}]_j^+.$$
The middle term vanishes. Now,$$[M_j^-,B_{k\a l}]_j^+=[M_j^-,B_{k\a l}]_j-
[M_j^-,B_{k\a l}]_j^-=[M_j^-,B_{k\a l}]_k^-$$ by the same reason: a global
function vanishing at infinity is a sum of its principal parts. The obtained
term cancels with the remaining one.

(2) $k=j$. $$\p^*\p_{k\a l}\hw_k-\p_{k\a l}\p^*\hw_k=-\p^*R_k^+\hw_k-\p_{k\a l}
(-M_k^++\sum_{i\neq k}M_i^-)\hw_k$$ $$=\{[M_k^+-\sum_{i\neq k}M_i^-,R_{k\a
l}]_k^++[R_{k\a l}^+,M_k^+-\s_{i\neq k}M_i^-]-[M_k,B_{k\a l}]_k^++\s_{i\neq k}
[M_i,B_k]_i^-\}\hw_k.$$Three terms cancel: $$[M_k^+,R_{k\a l}]^++[R_{k\a l}^+,
M_k^+]-[M_k,R_{k\a l}^-]^+=[M_k^+,R_{k\a l}]^+-[M_k^+,R_{k\a l}]^+=0.$$ The
remaining terms are $$-[M_i^-,R_{k\a l}]_k^+-[R_{k\a l}^+,M_i^-]_k+
[M_i,R_{k\a l}^-]_i^-=-[M_i^-,R_{k\a l}^-]_k^+-[R_{k\a l}^+,M_i^-]_k^-+[M_
i^-,R_{k\a l}^-]_i^-$$ The middle term vanishes and $[M_i^-,R_{k\a l}^-]_k^+=[M
_i^-,R_{k\a l}^-]_i^-$ by the same reason as it was in the first part, and the
whole expression vanishes. $\Box$ \\

Now $t^*$ is a new variable, independent of all $t_{k\a l}$. This implies
$$\p^*w_k=\p^*(\hw_k\exp\xi_k)=(\p^*\hw_k)\exp\xi_k=(-M_k^++\s_{i\neq k}M_i^-)
w_k.$$ The dressing formula $\p_{k\a l}-B_{k\a l}=w_i\p_{k\a l}w_i^{-1}$
implies that $$\p^*B_{k\a l}=[-M_k^++\sum_{i\neq k}M_i^-,-\p_{k\a l}+B_{k\a l}]
$$ Taking into account (7), this yields $$\p^*B_{k\a l}=[-\p_z+\s_iM_i^-,
-\p_{k\a l}+B_{k\a l}].$$ Take a linear combination with coefficients $\l_{k\a
l}$ of these equations. Then (see Eq.(6))$$\p^*U=[-\p_z+\s_iM_i^-,\p+U].
\eqno{(9)}$$

{\bf Definition.} {\sl A string equation is a condition that $U$ does not
depend on $t^*$. This yields} $$[\s_iM_i^-,\p+U]=\p_zU.\eqno{(10)}$$

In terms of $\hw_j$, the string equation has a form $$\p_z\hw_j=\s_{\a l}t_{j\a
l}lR_{j\a(l+1)}^+\hw_j-\s_{i\neq j}\s_{\a l}t_{i\a l}lR_{i\a(l+1)}^-\hw_j,
\eqno{(11)}$$ and in terms of $w_j$: $$\p_zw_j=-\s_{i\a l}t_{i\a
l}lR_{i\a(l+1)}
^-w_j.\eqno(12)$$ We have found here an additional symmetry. A problem arises:
are there other symmetries, as it is the case with KP, where there is a
infinite series of them labeled with two integer indices? We have no answer
to this question yet.

{\bf 5. Grassmannian.} We use the following notations: $C_k$ are
disjoint circles around the fixed
points $a_k$, $k=1,...,m$, and $\Omega$ is the part of the Riemann sphere
outside all the circles; $H_k$ are Hilbert spaces of vector-functions
$\bof(z)\in {\bf C}^n$ on the circles, subspaces $H_k^+$ consist of functions
on $C_k$ which can be expanded in non-negative powers of $z-a_k$, and $H_k^-$
contain expansions in negative powers. Now, $H=\bigoplus_k H_k$, and $H^+$
consists of $\{\bof_k\}$ such that $\bof_k\in H_k^+$ under an additional
constraint $\sum_k\bof_k(a_k)=0$.

Finally, let $H^*$ consist of $\bof=\{\bof_k\}\in H$ such that $\bof_k$ are
boundary values on the circles $C_k$ of a holomorphic vector-function in the
domain $\Omega$; this function will be denoted by the same letter $\bof$. It is
easy to see that $H=H^*\oplus H^+$. Indeed, let $\bof=\{\bof_k\}\in H$ be an
arbitrary element. Then each $\bof_k$ can be decomposed
into $\bof_k=\bof_k^-+\bof_k^+$. Elements $\bof_k^-$ are holomorphic outside
the corresponding circles $C_k$, and elements $\bof_k^+$ are holomorphic
inside the corresponding circles. Now,
$$\bof_k=(\sum_i\bof_i^-+\bc)+(\bg_k-\bc)
$$ where $\bg_k=\bof_k^+-\sum_{i\neq k}\bof_i^-$ and $\bc=m^{-1}\sum_k\bg_k(a_k
)$. We have, $\{\sum_i\bof_i^-+\bc\}\in H^*$ and $\bg_k-\bc\in H^+$. The
decomposition is unique. Let $P^*$ be the projector $P^*:H\rarr H^*$.\\

{\bf Definition.} {\sl An element of the Grassmannian, $W\in$Gr, is a subspace
of $H$ with the following properties: i) the projection $P^*:H\rarr H^*$
restricted to $W$ is a bijection, and ii)} $(z-a_1)^{-1}W=(z-a_2)^{-1}W=...
=(z-a_m)^{-1}W\subset W$.\\

We think about vectors as vector-rows. A matrix is said to belong to $W$ if so
do all its rows.

One can consider the following transformation of the
Grassmannian. If $\bof=\{\bof_k\}\in W$ then $\bof\exp\xi=\{\bof_k\exp\xi_k\}$
where $\xi_k=\sum_{l=0}^\infty\sum_{\a=1}^nt_{k\a l}E_{\a}(z-a_k)^{-l}$, $\xi=
\{\xi_k\}$. The set of all $\bof\exp\xi$ is called $W\exp\xi$. For almost all
$t_{k\a l}$ the subspace $W\exp\xi$ is a new element of the Grassmannian. \\

{\bf Definition.} {\sl A Grassmannian pre-Baker function, corresponding to an
element of the Grassmannian $W$, is a matrix-function $w\in W$ such that $$P^*w
\exp(-\xi)=c$$ where $c$ is a constant in $z$ (however, it can depend on
variables $t$).} \\

An element $W\in$Gr is invariant with respect to multiplication on the left by
a matrix constant in $z$, since this leads to linear combinations of rows. The
projector $P^*$ commutes with this multiplications. Therefore, if $w$ is a
pre-Baker function then so is a gauge-equivalent function $cw$ where $c$ is any
matrix independent of $z$. All pre-Baker functions can be expressed in this way
in terms of one of them, e.g., corresponding to $c=I$ (the normalized pre-Baker
function).

Let $w=\{w_k\}$ be a pre-Baker function and $\hw_k=w_k\exp(-\xi_k)$. This means
that $\hw_k$ has a form $c+w_{k,0}+w_{k,1}(z-a_k)+...$ and $\sum w_{k,0}=0
$. For a normalized function $c=I$. Thus, the normalized pre-Baker
function has expansions  $$w_k=(I+w_{k,0}+w_{k,1}(z-a_k)+...)\exp\xi_k\in W,~
k=1,...,m;~\sum_kw_{k,0}=0.$$ These equalities are equivalent to the
definition of the normalized pre-Baker function.
Let $w$ be a pre-Baker function. We have $$\p_{k\a l}w_i-R_{k\a l}^-w_
i=(\p_{k\a l}\hw_i+\d_{ki}\hw_iE_{\a}(z-a_k)^{-l}-R_{k\a l}^-\hw
_i)\exp\xi_i$$
$$=\left\{\begin{array}{ll}(\p_{k\a l}\hw_i-R_{k\a l}^-\hw_i)
\exp\xi_i,&i\neq k\\(\p_{k\a l}\hw_i+R_{k\a l}^+\hw_i)\exp\xi_i,
&i=k.\end{array}\right.\eqno{(13)}$$

{\bf Definition.} {\sl A pre-Baker function is called a Baker function if for
every $(k\a l)$ a relation $\{\p_{k\a l}w_i\}\in (z-a_j)^{-1}W$ holds (in fact,
this subspace does not depend on $j$, see the definition of the
Grassmannian).}\\

{\bf Proposition.} {\sl A Grassmannian Baker function is a formal Baker
function of the hierarchy, in the sense of Sect.3. Thus, a
solution of the hierarchy equations is related to any Grassmannian Baker
function.}\\

{\it Proof.} The left-hand side of (13), $\{\p_{k\a l}w_i-R_{k\a l}^-w_i\}$,
belongs to $(z-a_k)^{-1}W$. Therefore, the expression in parentheses in the
right-hand side, let it be $g_i$, is in $(z-a_k)^{-1}W\exp(-\xi)$. Then
$\{(z-a_j)g_i(z)\}\in W\exp(-\xi)$ for every $j$. On the other hand, this is
an element of $H^+$ plus, maybe, a constant. Let $\g_j$ be arbitrary matrices.
Then $\{\s_j\g_j(z-a_j)g_i(z)\}\in W\exp(-\xi)$. It is easy to see that
choosing
matrices $\g_j$ one can achieve that $\s_i\s_j\g_j(a_i-a_j)g_i(a_i)=0$. Then
$\{\s_j\g_j(z-a_j)g_i(z)\}$ is in $H^+$ and, therefore, in $W\exp(-\xi)\cap H^+
$. This implies that $\{\s_j\g_j(z-a_j)g_i(z)\}=0$ and $g_i=0$. The latter
means that $\hw_i$ satisfy the equations of the hierarchy. $\Box$\\

{\bf 6. An example: soliton-type solution.} Soliton solution were given first
in the original paper by Zakharov and Shabat [2]. Here a
different construction is presented. This approach is known for a long time:
Manin [9], Date [10], maybe even earlier, since Manin referres to Drinfeld. We
just tried here to do this in the possibly most general form and connect it to
the Grassmannian and to a $\tau$-function. In frameworks of the Grassmannian
theory this construction looks very natural.

One must start with a specification of an element $W\in$Gr. Consider a linear
space $H^*(D)$ of meromorphic vector-functions $\bof$ in the domain $\Omega$
with a fixed divisor $D$ of simple poles $b_j$,
where $j=1,...,N$. Collections of boundary values of these functions, $\{\bof_k
\}$, on the circles $C_k$ will be denoted by the same letter $\bof$, and the
linear space of them by the same symbol $H^*(D)$. This will not lead to any
ambiguity.

Now, let $W\subset H^*(D)$ be a subset of meromorphic functions in $H^*(D)$
satisfying $Nn$ conditions ${\bf v}(\mu_i)\cdot{\bf \eta}_i=0$
where $i=1,..,Nn$, $\mu_i\in\Omega$ are arbitrary points, and ${\bf\eta}_i$ are
given vector-columns. Collections of their boundary values are symbolized by
the same letter $W$, and this, generically, is an element of the Grassmannian.
The property ii) of the definition of the Grassmannian is self-evident. Notice
that $(z-a_j)^{-1}W$ consists of those elements which are boundary values of
meromorphic functions vanishing at infinity.

Let us prove that the property i) is also satisfied.
We have to prove that if there is an element $\bof=\{\bof_k\}\in H^*$ then a
unique element $\bg=\{\bg_k\}\in W$ can be found such that $\bh=\bof-\bg=\{\bof
_k-\bg_k\}=\{\bh_k\}$ is in $H^+$, i.e., its analytical prolongation inside
every circle $C_k$ exists, being $\sum_k\bh_k(a_k)=0$.
Given $\bof_k$ are boundary values of a holomorphic in $\Omega$ function
$\bof$. Let $\bg=\bof+\bb_0+\sum_{j=1}^N\bb_j(z-b_j)^{-1}$ where $\bb_j$ are
vectors that have to be found, in all $(N+1)n$ unknown components. First
of all we require that $\bg\in W$ which is equivalent to $Nn$ scalar equations
$\bg(\mu_i)\cdot{\bf\eta}_i=0$. Then we impose one more constraint
$\sum_{k=1}^m(\bb_0+\sum_{j=1}^N\bb_j(a_k-b_j)^{-1})=0$ which gives $n$ more
equations for the unknown coefficients, in all $(N+1)n$ equations.
Generically, this system can be solved uniquely. After that,
boundary values of $\bb_0+\sum_j\bb_j(z-b_j)^{-1}$, call them $\bh_k$, belong
to $H^+$, and $\bg_k=\bof_k+\bh_k$ that implies $P^*\bg=\bof$, as required.

We are looking for a Baker function in the form $$w=\Phi(z)\exp\s_k\xi_k=(I+
\s_{j=1}^NB_j(z-b_j)^{-1})\exp\s_k\xi_k.$$ Here $B_j$ are
matrices. Then $\hw_k=(I+\s_{j=1}^NB_j(z-b_j)^{-1})\exp\s_{i\neq k}\xi_i$.
We have the following equations for elements of the matrices $B_j$
$$\Phi(\mu_i)\exp\sum_k\xi_k(\mu_i)\cdot{\bf\eta}_i=0,~~i=
1,...,Nn$$or, in coordinates, $$\sum_{\b=1}^n\Phi_{\a\b}(\mu_i)y_{\b i}=0,~
\a=1,...,n,~i=1,...,Nn \eqno{(14)}$$ where $$y_{\b i}=\exp[\sum_{k=1}^m\sum_
{l=1}^\infty t_{k\b l}(\mu_i-a_k)^{-l}]\eta_{i\b}.$$

Matrices $B_j$ are determined by this equation uniquely. We have chosen a gauge
in which $w_i$ are boundary values of a function constant at infinity,
therefore $\p_{k\a l}w_i$ are boundary values of a function vanishing at
infinity and $\{\p_{k\a l}w_i\}\in (z-a_j)^{-1}W$, i.e. this is a Baker
function.

{\bf Proposition.} {\sl The solution to the system (14) is given by the
formula
$$\Phi_{\a\b}={1\over\Delta}
\left|\begin{array}{ccccccc}\delta_{\a\b}&|&y_{\a1}&.&.&.&y_{\a,Nn}\\
-&-&-&-&-&-&-\\(z-b_1)^{-1}\d_{1\b}
&|&(\mu_1-b_1)^{-1}y_{11}&.&.&.&(\mu_{Nn}-b_1)^{-1}
y_{1,Nn}\\.&|&.&
.&.&.&.\\(z-b_1)^{-1}\d_{n\b}&|&(\mu_1-b_1)^{-1}y_{n1}&.&.&.&(\mu_{Nn}-b_1)^
{-1}y_{n,Nn}\\-&-&-&-&-&-&-\\.&.&.&.&.&.&.\\-&-&-&-&-&-&-\\(z-b_N)^{-1}\d_{1\b}
&|&(\mu_1-b_N)^{-1}y_{11}&.&.&.&(\mu_{Nn}-b_N)^{-1}y_{1,Nn}
\\.&|&.&.&.&.&.\\(z-b_N)^{-1}\d_{n\b}&|&(\mu_1-b_N)^{-1}
y_{n1}&.&.&.&(\mu_{Nn}-b_N)^{-1}y_{n,Nn}\end{array}\right| $$
Here $\Delta$ is the cofactor of the element $\delta_{\a\b}$}.\\

(The structure of the determinant is the following. It has $Nn+1$ rows and
columns. All the rows except the first one can be parted into $N$ groups, $n$
rows in each of them. The rows, except the first one, can be labeled by
$j,\gamma$ where $j=1,...,N$ and $\gamma=1,...,n$. The columns, except the
first one are labeled by $i=1,...,Nn$. The non-zero entries of the first
column are on the $(j,\b)$ places, i.e., on the $\b$th place in each group, and
also the upper left element, if $\a=\b$.)

{\em Proof.} Left-hand side of Eq.(14) is represented by a determinant where
the first column coincides with the $i$th, hence it vanishes. Taking into
account the division by $\Delta$, we see that $\F$ has a desired form
$I+\s_{j=1}^NB_j(z-b_j)^{-1}$. $\Box$\\

{\bf 7. Expression of the Baker function in terms of $\tau$-functions for
solitons.} The next very natural topic in this context would be a
$\tau$-function. According to the common definition of that, introduced by
Sato et al., see [1], we could expect a relation between the Baker
and the $\tau$ functions something like
$$w_{k,\a\b}={\tau_{k,\a\b}(t_{k_1\g l}-\d_{kk_1}\d_{\g\b}l^{-1}(z-a_k)^l)\over
\tau(t)}\exp\sum_i\xi_i(z).\eqno{(15)}$$ We do not give here a general
definition of a $\tau$-function and restrict ourselves to the soliton-type
solutions. We show that in this case the formula (15) can be written for
some $\tau$'s, indeed.

In order to obtain $\hw_k$, one has to multiply $\F$ by $\exp\s_{k_1\neq k}\xi_
{k_1}(z)$ and expand it in powers of $z-a_k$. There are two cases, i) $\a=\b$
and ii) $\a\neq\b$.

i) Let us transform the determinant adding the first row multiplied by $-(z-b_j
)^{-1}$ to all $j\b$th rows, $j=1,...,N$ , i.e., annul all elements of the
first column except the first one. Expanding along the first column, we get
that
$\Phi_{\b\b}=\prod_{j=1}^N(z-b_j)^{-1}$ multiplied by a $N\times N$
determinant with the entries: $(z-\mu_i)(\mu_i-b_j)^{-1}y_{\b i}$ on the $(\b j
,i)$ place and $(\mu_i-b_j)^{-1}y_{\gamma i}$ on the $(\gamma j,i)$ place for
$\gamma\neq\b$. An obvious identity $$\mu_i-z=(\mu_i-a_k)\exp[-\s_{l=1}^\infty
{1\over l}({z-a_k\over\mu_i-a_k})^l]$$ implies $$\hw_{k,\b\b}=
\prod_{j=1}^N(z-b_j)^{-1}\Delta^{-1}\det(\tilde{T}_{\gamma j,i}^{k\b\b})\exp
\sum_{k_1\neq k}\sum_{l=1}^\infty t_{k_1\b l}(z-a_{k_1})^{-l},$$where
$(\tilde{T}_{\gamma j,i}^{k\b\b})$ with fixed $k$ and $\b$ is a $Nn\times Nn$
matrix, $\g j$ is a number of a row and $i$ that of a column,
$$\tilde{T}_{\gamma j,i}^{k\b\b}=\left\{\begin{array}
{cc}-{\mu_i-a_k\over\mu_i-b_j}\tilde{y}_{\b i}^{k\b},&{\rm if}~\gamma=\b\\
{1\over\mu_i-b_j}y_{\gamma i},&{\rm otherwise}\end{array}\right.$$ We denoted
$$\tilde{y}_{\gamma i}^{k\b}=\exp[\sum_{k_1=1}^m\sum_{l=1}^\infty (t_{k_1\gamma
l}-\d_{k_1,k}\d_{\g\b}{1\over l}(z-a_k)^l)(\mu_i-a_{k_1})^{-l}]\eta_{i\gamma}.
$$ The factor $\prod_{j=1}^N(z-b_j)^{-1}$ does not play any role in the
dressing
formula (4), it just cancels out. Thus, except for the factor $\exp\sum_{k_1}
\sum_{l=1}^\infty t_{k_1\b l}(z-a_{k_1})^{-l}$ the whole dependence on $z$ is
in modified time variables $t_{k_1\gamma l}\mapsto t_{k_1\gamma l}-\delta_{k_1,
k}\delta_{\g\b}l^{-1}(z-a_k)^l$. This is just what we need in order to obtain
(15). Thus, a $\tau$ function for a matrix element
$w_{k,\b\b}$ is $\tau_{k,\b\b}=\det T^{k\b\b}$ where $$T_{\gamma j,i}^{k\b\b}=
\left\{\begin{array}{cc}-{\mu_i-a_k\over\mu_i-b_j}y_{\b i},&{\rm
if}~\gamma=\b\\
{1\over\mu_i-b_j}y_{\gamma i},&{\rm otherwise}\end{array}\right.$$

ii) Now the element $\Phi_{\a\b}$ with $\a\neq\b$. The $(1\b)$ row multiplied
by $(z-b_1)(z-b_j)^{-1}$ must be subtracted from the $(j\b)$ row, for all
$j=1,...,N$. We have $$\hw_{k,\a\b}=\prod_{j=1}^N(z-b_j)^{-1}\Delta^
{-1}\det(\tilde{T}_{\gamma j,i}^{k\a\b})\exp\sum_{k_1\neq
k}\sum_{l=1}^\infty t_{k_1\b l}(z-a_{k_1})^{-l},$$where$$\tilde{T}_{\gamma j,i}
^{k\a\b}=\left\{\begin{array}{cc}y_{\a i},&{\rm if}~j=1,\g=\b\\
{(\mu_i-a_k)(b_j-b_1)\over(\mu_i-b_j)(\mu_i-b_1)}\tilde{y}_{\b i}^{k\b},&{\rm
if}~j>1,\gamma=\b\\{1\over\mu_i-b_j}y_{\gamma i},&{\rm otherwise}\end{array}
\right.$$ The determinant $\det T^{k\a\b}$ is a $\tau$-function for $w_{k,\a
\b}$, i.e., $\tau_{k,\a\b}$ where $$T_{\gamma j,i}^{k\a\b}=\left\{\begin{array}
{cc}y_{\a i},&{\rm if}~j=1,\g=\b\\{(\mu_i-a_k)(b_j-b_1)\over(\mu_i-b_j)(\mu_i-
b_1)}y_{\b i},&{\rm if}~j>1,\gamma=\b\\{1\over\mu_i-b_j}y_{\gamma i},&{\rm
otherwise}\end{array}\right.$$

{\em In this particular example we obtained the following fact. There are
matrix-functions $\tau_k(t)$ such that Eq.(15) holds, being} $\tau(t)=\Delta$.
\\

It would be interesting to give a general definition of the $\tau$-function in
terms of the Grassmannian similar to that given in a single-pole case in [4].\\

{\bf References.}\\

\noindent 1. Date, E., Jimbo, M., Kashiwara, M., and Miwa, T.: Transformation
groups for soliton equations, in Jimbo and Miwa (eds.) Non-linear integrable
systems, classical and quantum theory, Proc. RIMS symposium, Singapore (1983)\\
2. Zakharov, V. E., and Shabat, A. B.: Integration of nonlinear equations of
mathematical physics by the method of inverse scattering, Funct. Anal. Appl.,
13, No3, 13-22 (1979)

Zakharov, V. E., Manakov, S. V., Novikov, S.P., and Pitajevski, L. P.:
Theory of solitons (1980)\\
3. Dickey, L. A.: Soliton equations and Hamiltonian systems,
Advanced Series in Math. Physics, vol. 12, World Scientific, (1991)\\
4. Dickey, L. A.: On Segal-Wilson's definition of the $\tau$-function and
hierarchies AKNS-D and mcKP, to appear in Proc. of the Lumini Conf. on
Integrable Systems, July (1991).\\
5. Chen, H. H., Lee, Y. C., and Lin, J. F.: On a new hierarchy of symmetries
for
the Kadomtsev-Petviashvili Equation, Physica D, 9D, No3, 439-445 (1983)

Orlov, A. Yu., and Shulman, E. I.: Additional symmetries for integrable
equations and conformal algebra representations, Lett. Math. Phys., 12, 171-179
(1986)\\
6. Dickey, L. A.: Additional symmetries of KP, Grassmannian, and the string
equation, I, II, Preprints hep-th 9204092, 9210155 (1992), to be printed in
Modern Phys. Letters \\
7. Segal, G., and Wilson, G.: Loop groups and equations of KdV-type, Publ.
Math. IHES, 63, 1-64 (1985)\\
8. Harnad, J., and Wisse, M. A., Moment maps to loop algebras, classical
R-algebras and integrable systems, preprint, hep-th 9301104 (1993)\\
9. Manin, Yu.: Matrix solitons and vector bundles over curves with
singularities, Funct. Anal. Appl., 12, No4, 53-67 (1978)\\
10. Date, E.: On a direct method of constructing multi-soliton solutions,
Proc. Japan Acad., A55, 27-30 (1979)\\

\end{document}